\documentclass[twocolumn,showpacs,titlepage,aps,prd,letter,
tightenlines,amsmath,byrevtex,nofootinbib,floatfix]{revtex4}

\usepackage{graphicx}

\begin{document}

\preprint{hep-ph/????}
\title{Galactic Neutrino Communication} 

\author{John G. Learned$^{1}$}
\email{jgl@phys.hawaii.edu}
\author{Sandip Pakvasa$^{1}$}
\email{pakvasa@phys.hawaii.edu}
\author{A. Zee$^{2}$}
\email{zee@kitp.ucsb.edu}
\affiliation{
$^1$Department of Physics and Astronomy, University of Hawaii, 2505 Correa Road, Honolulu, Hawaii 96822 U.S.A.\\
$^2$Kavli Institute for Theoretical Physics, University of California, Santa Barbara, California 93106 U.S.A. \\
}

\date{\today}

\vglue 1.6cm
\begin{abstract}

We examine the possibility to employ neutrinos to communicate within 
the galaxy. We discuss various issues associated with transmission and 
reception, and suggest that the resonant neutrino energy near 6.3 PeV 
may be most appropriate. In one scheme we propose to make $Z^o$ 
particles in an overtaking $e^+ - e^-$ collider such that the resulting 
decay neutrinos are near the $W^-$ resonance on electrons in the 
laboratory. Information is encoded via time structure of the beam. In 
another scheme we propose to use a $30 PeV$ pion accelerator to create 
neutrino or anti-neutrino beams.  The latter encodes information via 
the beam CP state as well as timing.  Moreover the latter beam requires 
far less power, and can be accomplished with presently foreseeable 
technology.  Such signals from an advanced civilization, should they 
exist, will be eminently detectable in existing neutrino detectors.

\end{abstract}


\maketitle

\section{Introduction: Neutrinos for galactic communication}
\label{introduction}

The search for extraterrestrial intelligence (SETI) has now gone on 
for decades without detecting a signal. Thus far, the search has 
presumed that the transmission would be in photons within radio or 
optical bands. In this note, we suggest that it may be more 
sensible to search for a possible signal in neutrinos. We will 
discuss some of the physics issues related to both transmission and 
reception.

Why would one want to employ the notoriously difficult-to-observe 
neutrino to communicate? Some reasons are obvious.

i) The obscuration/opaqueness of material between the source and 
detector makes photons less useful within the galactic plane and 
particularly for lines-of-sight anywhere near the galactic center.  
Neutrinos arrive almost without attenuation from any source 
direction (though at the energies we suggest below, they are 
attenuated through the earth).

ii) Neutrinos are rare, and from a given direction are all but 
negligible (particularly at high energies as we discuss below).  
For photons the signal/noise (S/N) problem is not at all 
negligible in any band and is worse for all directions where 
galactic civilizations may reside (presumably in the galactic 
plane).  For neutrinos the band can be essentially noise free.

iii) Even when photons are not completely blocked, their scattering 
introduces jitter in arrival time as well as direction.  As we 
discuss below, the reasonable encoding for maximally 
energy-efficient information transfer may employ the time interval 
between quanta.  Moreover, in one scheme, data may be encoded via 
the use of neutrinos versus anti-neutrinos.

There have been early general proposals for the use of neutrinos for 
interstellar communication and for SETI \cite{pasach}. There were also 
more recent and more specific proposals about the use of neutrinos in 
SETI \cite{lpst}, \cite{silagadze}. In this note we propose a new way 
of using neutrinos for communication by ETI and point out that 
existing and future facilities will be able to look for these signals 
without any need for new construction or expense.

\section{What neutrino energies are most suitable?}
\label{neutrino energies}

What neutrino energies are best suited for galactic communication? 

First, let us consider relatively low energy neutrinos, those 
typical of nuclear, solar and supernovae (SN) processes.  One 
strike against this energy region, roughly up to around 10 MeV (40 
MeV for SN) is that such neutrinos are produced in abundance 
naturally, so there is some signal-to-noise barrier to overcome. If 
one thinks of the natural emission from radioactive decays, from 
stars and from supernovae across the universe, this region is not 
so attractive since these sources are enormously powerful and one 
must compete against them (The ``noise" is small, but the inherent 
signal-to-noise ratio S/N is also small for any imaginable source.)  
Moreover, directionality at the lower energies is at best 
difficult.  Also, S/N aside, at low energies the neutrino-nucleon 
interaction cross-section is dauntingly small, of order 
$10^{-42}~cm^2$.  (For the same reason of vanishingly small 
cross-sections, we dismiss further discussion of even lower 
neutrino energies.) We have also considered the possibility of 
employing resonant nuclear energies, but we have not identified any 
viable mechanism which will beat the problems of inherent S/N and 
low cross sections.

For these reasons we are driven to consider higher energies. The 
neutrino cross section grows with energy linearly until the 100 TeV 
energy range, and then logarithmically.  The (dominant) terrestrial 
neutrino backgrounds fall rapidly, one power in energy more steeply 
than the cosmic ray spectrum. For a beam, there are further gains 
of order $E^2$ due to the solid angle of the beam, and typically 
one would gain altogether by a factor of $>E^3$ in signal, and 
$>E^6$ in S/N on a per particle basis.

The energy chosen should be such that it would be clear at once that it 
is an artificial source such as ETI and not some random background. One 
choice is to make $Z^o$'s at rest as was proposed earlier \cite{lpst}. 
Then the neutrinos from such a source have energies of exactly $m_Z$/2, 
about 45 GeV, and are easily identifiable as due to $Z^o$ decay: there 
are no natural sources of $\nu$'s of this precise energy. In this case 
the neutrinos are emitted in a spherically symmetrical manner, and 
because of that the power requirements for galactic distances, reach 
the scale of total solar power (as estimated there) to obtain a 
significant counting rate. Of course one might argue that this is not 
``our" problem, but one to be solved by the postulated advanced 
civilization with technology we cannot yet imagine.  But resorting to 
harnessing (Dyson) stars certainly moves the potentiality of such 
communication to the distant future, if indeed such is ever practical 
for a civilization. It was further proposed there that the detection 
process employed electrons accelerated such that the incoming 
electron-anti-neutrino interact at the Glashow resonance\cite{glashow} 
i.e. producing a W boson on-shell (an electron energy $E_e = {m_W}^2 /4 
m_e$, about 35.5 GeV) with a large cross-section.

We presume that the ETIs, though in our galaxy, are remote. Even if an 
ETI has been observing us, it may be a long while (timescale of 
thousands of years) before they would send us an introductory message. 
So if they want to send a message in advance, saying hello and welcome 
to the galactic network, they are going to have to speculate about 
when to bother to transmit. From the jittering of advances in 
speciation, with the great die offs, it seems clear that evolution is 
a stochastic process, with fluctuations on a timescale of many 
millions of years.  The evolution of technology may fluctuate over a 
timescale of thousands of years, as exemplified by the long periods of 
lack of technological progress in post-Roman Europe, China and India.  
One must reason that no useful prediction could be made as to when the 
industrial revolution would take off and high technology would arise.  
Thus the ETI would have to be transmitting speculatively over a long 
period.

Communication may be initiated for many hypothetical reasons.  For 
example, to simply welcome a new society to the galactic club, or 
to warn of dangers from within or without, or for reasons we cannot 
now guess. Indeed, it seems best not to speculate on motives for 
transmission at this point but to focus on physics issues.  In any 
event, it seems that there might be two stages, the first being to 
simply get the attention of the recipient and the second to send 
information.  One might also imagine ETI sending information to 
military outposts via secure neutrino beam, on a known schedule and 
in a small solid angle, and we happen to be in the transmission 
path, but the probability of our intercepting such is evidently 
negligible.

Given light transmission time over galactic dimensions, compared to 
evolution time in our world, we cannot foresee much of a dialogue, 
but that is of course also true for electromagnetic communication.  
Note, for example, that signals leaving the center of our galaxy at 
the time of the first human settlements would not arrive for 
another 14 millenia. Monologues are what we can anticipate at best. 
That is the case unless there are means to beat the speed of light 
via wormholes or extra dimensions, but such do not now appear to be 
viable within established physics.

It has been argued persuasively\cite{rose} that transmission of large 
amounts of data via radio or light are very inefficient compared to the 
deposition of artifacts, snail mail as it were. Artifacts can hold huge 
amounts of data, and need only be sent once or very seldom to promising 
star systems and left to be discovered later.  We thus imagine that 
there is not a great need for a high data rate channel in the galaxy, 
but perhaps only as stated above, for some minimal information of 
overwhelming importance, which might include instructions on where to 
find the artifact.

\section{Directional transmission and detection via resonant neutrinos}
\label{concept}

Instead of omnidirectional broadcast, let us next consider the 
possibility of sending out focused neutrino beams in a specified 
direction, aimed at promising star systems.

Sending out a focused beam has the advantage of not being seen by all, 
perhaps a worthy security measure.  Many have speculated on the danger 
of attracting unwanted attention by potentially aggressive species 
(which conceivably could wish to take over a nice proven habitable 
planet, or even enslave the inhabitants of such a planet --- the 
subject of much science fiction but not obviously a wrong 
presumption.)  Indeed, some have suggested that our civilization 
should consider measures to keep our galactic visibility to a minimum 
for just this reason.  Hence transmitting to an unknown new society 
has risks, particularly since with a beam one may reveal the location 
of the transmitting entities.  If this is indeed a real danger, 
perhaps an advanced civilization would employ a transmitting station, 
a lighthouse, at some remove from their home. In view of this, perhaps 
an advanced civilization may be more inclined to transmit to a newly 
technically emergent society (TES) such as ours. Such action might 
offer the rewards of heading off more undesirable consequences 
possible in the longer term, when a TES is not in a position to be too 
territorially acquisitive. Or perhaps the ETI would simply like to 
initiate trade as when Europeans first visited China and India.  
However, our own history gives no precedent in terms of communication 
prior to contact, and indeed that history is a bit frightening, since 
first contacts soon led to exploitation.

Let us assume, moreover that the ETI will guess that a civilization 
ready to hear their messages will have developed to the point of 
constructing large neutrino detectors in the process of studying 
neutrinos and attempting to begin neutrino astronomy.  It is hard to 
justify from some future viewpoint, but we see now that a high energy 
neutrino detector of the scale of 1 km is reasonable (IceCube under 
construction at the South Pole, and NESTOR, ANTARES and NEMO and an 
expanded Lake Baikal detector are all proposed or under 
development\cite{lm}).  So, for discussion purposes we shall assume the 
ETI will aim at communicating with a detector of cross-section on the 
order of $1 km^2$. 

\begin{figure}[htbp]
\begin{center}
\includegraphics[width=0.5\textwidth]{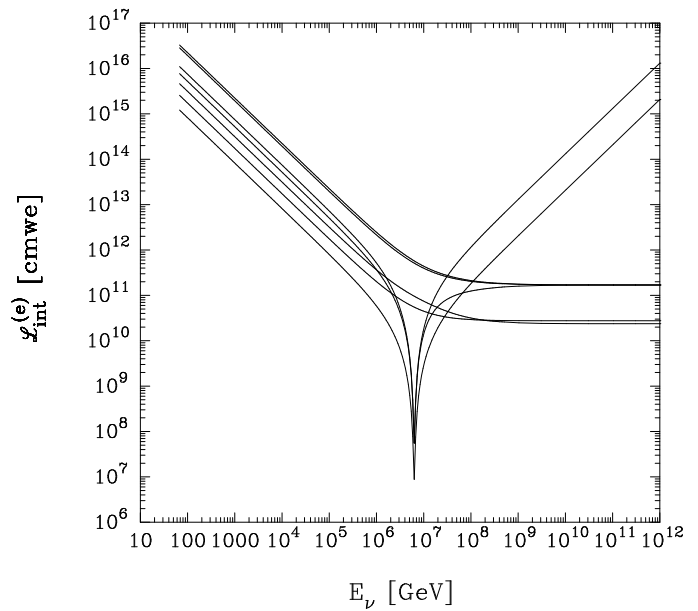}
\end{center}
\caption{Electron and muon (and tau same as muon) neutrino and 
anti-neutrino interaction lengths on target electrons versus 
neutrino energy, including Glashow resonance. From \cite{gqrs98}. }
\label{fig:crossection}
\end{figure}

Suppose we make a beam of electron-anti-neutrinos, and get them to a 
very high energy, to be precise $E_G = {M_W}^2/2 m_e = 6.3 PeV$.  Then 
these can be sent in some chosen directions to be received by observers 
who can employ a detector seeking the reaction $\bar{\nu_e} + e^- 
\rightarrow W^-$ at resonance, the so-called Glashow resonance, as
illustrated in Figure \ref{fig:crossection}. The production and decay
of a $W^-$ into a shower provides a unique signature; given more than
one such event from a given direction the source would be immediately
known to be due to an ETI as there are no natural sources of 6.3 PeV
$\bar{\nu_e}$'s. To contrast with the proposal made in 1994\cite{lpst},
here we boost the initial beam rather than the electrons in the
detector; in both one employs the Glashow resonance and its high
cross-section. The range of such a resonant neutrino in water is about
100 km, so that these neutrinos would penetrate to the deepest detectors
on earth, but would be attenuated in arrival directions below the
horizon. This also implies that the detection fraction in a $1~km^3$
detector would be about 1\% of the traversing neutrinos.

An efficient mechanism for such a beam generation would be to collide 
electron-positron beams at a center-of-mass energy at the $Z^o$ mass, 
but in a fast moving reference frame, so that the decay neutrinos 
would be at 6.3 PeV.  In this instance the electrons might be 
overtaking the positrons in the laboratory frame, so that the $Z^o$ is 
fast forward moving and the beam direction determined thereby.

The size of the region illuminated by such a beam from a distance 
of 1 kiloparsec (about 3,000 light years, $3 \times 10^{16}$ km) is 
about 3,000 AU across.  If we require a beam such that there would 
be at least 100 neutrinos per $km^2$ area, then the individual 
pulse would have to have around $10^{26}$ neutrinos(here we use an
opening angle in the Z decay of about $M_{Z}/E_{\nu}$).

Needless to say, the numbers we cite are illustrative only as we 
clearly could not anticipate all possible scenarios. For instance, the 
ETI may be relatively nearby, either on a extrasolar planetary base or 
in a space station, waiting for us to build a suitable neutrino 
telescope.

\section{Energy Costs}
\label{energy}

Accelerators are marvelously good at transforming electrical energy 
into beam energy, typically putting tens of percent of the wall-plug 
power into the beam.  One can imagine a total energy transfer from the 
delivered power to neutrinos of perhaps 0.1\%.  In contrast, radio is 
better, by perhaps one or two orders of magnitude. Lasers are currently 
not very efficient, but their technology is still evolving rapidly. 
One may thus argue that the cost of making a neutrino beam is not 
prohibitive in comparison to photon beams. Also, the radio beams have 
the potential disadvantage of sidelobes, which could be a strongly 
negative factor if the security issue is real or perceived to be so.

At an energy cost of order $10^{25}$ Joules per pulse (allowing for
accelerator efficiency and the fact that the Z decays into
neutrinos only 20\% of the time, and of that only 1/6 will be
electron anti-neutrinos), this is a huge energy output.  If the
pulses were fired once per second, the accelerator power would be
about 3\% of the solar luminosity. In fact, taking into account
the flat spectrum of the neutrinos from Z decay after the boost,
there is a further factor of $m_Z/\Gamma_Z$, which makes the required
power about equal to the solar luminosity. This is clearly not a task we 
can imagine carrying out on any projection of our present 
technology.

However, we do not know the methods that may be available to 
advanced civilizations to make a neutrino (or any other) beam.  
We have direct evidence in the $10^{20}~eV$ cosmic rays, the 
gamma ray bursts (GRBs), the micro-Quasars, and the amazingly 
collimated jets from active galactic nuclei (AGN), so that we 
might suspect that we do not yet understand some fundamental 
issues on particle acceleration. For example, how does one get 
an earth mass accelerated to a gamma of 1000 in a distance of a 
few light seconds, as has been inferred for gamma ray burst 
jets or ``cannonballs"?  There is also the possibility mentioned earlier
of employing ``Dyson'' stars. So, for present purposes, we shall 
assume that an ETI would find it affordable and worthwhile to 
expend such resources to communicate with our TES.

\section{Information Encoding via Pulse Timing}
\label{encoding}

What about encoding? Since we are talking about relatively rare 
events, it seems evident that the encoding information by relative 
timing of the neutrino pulses provides the only mechanism, much as 
the use of a simple Morse code in the early days of electromagnetic 
communication. Also, neutrinos are fermions and presumably encoding 
could only involve classical physics, rather than say some 
hypothetical analog of the laser.

Neutrino oscillations might permit some further encoding, but given 
the distances, oscillations are averaged out.  For neutrinos 
coming from a distance near that to the galactic center, the solar 
oscillations will have made about a million cycles.  If the beam 
energy were sharp to parts per million, then in fact one would have 
to worry about the phase of the cycle (earth could be in a null), 
but this seems not a problem.  In any event if only electron 
anti-neutrinos are detected (as we are considering here), then no 
information can be encoded in neutrino flavor.

Thus there would be some interval between pulses, which we interpret 
as some number of time increments long, and that is the message.  
What would be the natural time increment? One possibility would be the 
lepton associated 0.3 picosecond lifetime of the tauon.  This timing 
over a large detector is not presently practical, and would seem to be 
not possible in the foreseeable future. The muon lifetime of 2.2 
microseconds would be much easier and more practical.  The minimum 
time interval detectable with present detectors is of the order of 1 
nanosecond, and may reach 100 picoseconds in a few decades. With 
maximum resolution, if events were spaced apart by, say (arbitrarily) 
2200 sec on average, then this is $10^{9}$ intervals per pulse, or 
equivalent to ~30 bits, or an equivalent data rate of about 0.014 
baud, not so bad for interstellar communications! If we assume 
transmission with, say, three repetitions then this would be 143,000 
bits per year, quite a respectable amount of data.

Some thought should be given to how the ETI might encode data in a 
way which would be most simple to decode.  For example, coarse 
level timing might encode for the simplest messages helping to 
establish the link and achieve synchronization, with finer and 
finer detail encoding more and more complex data.

Finally, at the risk of speculating beyond what would be warranted 
at present, we might try to guess the content of a message from an 
advanced civilization. We could ask what we might say if we were in 
a position to start transmitting. As suggested in another context 
\cite{hz}, in light of the physicist's well documented and almost 
irresistible urge to publish, an advanced civilization might just 
want to announce that it has figured out how the universe ticks. A 
concise summary would be the gauge algebras of the three 
non-gravitating interactions suitably coded.

\begin{figure}[htbp]
\begin{center}
\includegraphics[width=0.5\textwidth]{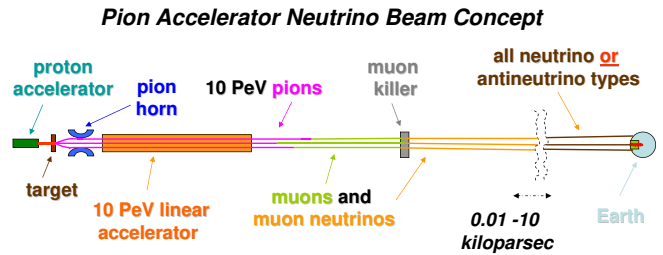}
\end{center}
\caption{Cartoon of muon neutrino production from a pion 
accelerator. Neutrinos mix and all flavors are detected, but only 
anti-neutrinos will have Glashow resonant events.} 
\label{fig:pion_beam}
\end{figure}

\section{An alternative with Neutrino SuperBeams}
\label{superbeams}

As this paper was being drafted, a preprint appeared \cite{silagadze} 
which proposed that communication from nearby stars might be conducted 
with neutrino beams such as will be inherent in a muon collider.  In 
present discussions of such, one would rapidly accelerate muons made 
from pion decays in the 1 GeV energy range (the pions made in 
collisions of a few megawatt proton beam of a few tens of GeV with a 
fixed target).  These muons would then circulate in order to study 
high energy muon pair collisions.  The muons decay and create a 
powerful (and even dangerous) neutrino beam. And, there has been much 
discussion of neutrino factories, using muons as discussed, but only 
keeping them in a race track (with long straight sides) shaped ring or 
a nearly triangular ring, with sides pointed at distant detector.  
Indeed such instruments seem to provide the path towards detailed 
measurements of neutrino mixing, and we think will be built in the 
foreseeable future.

We propose an alternative to this which seems to be even more 
interesting, as illustrated in the cartoon in Figure 
\ref{fig:pion_beam}.  Our idea is to accelerate the charged pions 
before they decay to muons, and to take them to energies in the range 
discussed previously, $\simeq 30 PeV$, so that the decays would make 
neutrinos in the 6.3 PeV energy range. The decay distance for pions of 
this energy will be about 0.5 million km, or about the distance to the 
moon.  The positively charged pions will decay into positive muons and 
muon neutrinos, and negative pions vice versa, with resulting 
anti-neutrinos. If one aims the beam at a relatively thin shield (rock) 
one will kill the muons (which radiate very copiously at this energy), 
but leave the neutrinos.  Hence in selecting whether positive or 
negative pions are accelerated, one may choose neutrinos or 
anti-neutrinos as the beam.

In case of $\pi^+$, only $\nu_{\mu}$'s are produced and from $\pi^-$ 
only $\bar{\nu_{\mu}}$'s are produces. As is well-known\cite{pakvasa} 
the averaged out oscillations, after a distance of about a few light 
days, convert these into a mixture of all three flavors in the 
proportion given by $\nu_e$:$\nu_{\mu}$:$\nu_{\tau}$ = 4:7:7, but 
keeping the particle/antiparticle nature as is.  Hence, since it is a 
matter of sign selection in the accelerator one can switch the neutrino 
beam between particles and antiparticles, switching on and off the 
Glashow resonance. There will be a constant signal from either beam due 
to the other charged and neutrino current reactions. Thus there will 
thus be different signatures for neutrinos and anti-neutrinos, and one 
may encode information in the nature of each pulse, without regard to 
timing.  In addition, timing can be employed as well, as discussed 
earlier for the resonant $Z^o$ scheme, and we can substantially boost 
the data transmission rate.

A strong further attraction of the pion scheme is that the maximum 
neutrino angle relative to the pion is $({m_\pi}^2 -{m_\mu}^2) / 2 
m_\pi / E_\nu$ so the beam is much narrower than the $Z^o$ beam and the 
target area is smaller by about a factor of about $10^7$. All 
transmitter powers involved are dramatically reduced so that we 
would be considering a total beam energy requirement per pulse of order 
one gigajoule.  This is on the order of the energy per pulse which is 
being discussed for near future controlled fusion reactions on Earth.  
Given such ``modest" power levels, one may imagine transmitting at 
rates higher than hypothesized for the resonant $Z^o$ method, perhaps 
one per second, as is easily foreseeable with present technology (a 
gigawatt of power, less than many present nuclear power stations).

The penalty for the tight beam however is that the ETI must know the 
precise planet they are targeting and know its ephemeris, since the 10 
million km beam spread (from 1 kpc) is less than 0.01 AU. If they have 
surveyed the stellar systems and singled out earth, then this would not 
seem a barrier as they would of necessity have determined the orbital 
parameters, and extrapolation of the earth's orbit over a few millenia 
should pose no problem.

The only thing we do not yet readily foresee in this scheme is how to 
accelerate the pions to this high energy ($30 PeV$).  Yet, a linear 
accelerator (in space) with a gradient of 100 GeV/m (already achieved 
for short distances in the laboratory) would require a length of 1,000 
km, which seems not wildly implausible for a future civilization.  
Also, while aiming with sufficient precision would certainly pose 
problems for us at present (at the level of 0.07 milliarcseconds), such 
a scale would be needed by the ETI to optically resolve the earth in 
any event.

\section{Concluding Remarks}
\label{conclusion}

We have outlined a method for intragalactic communication via directed 
6.3 PeV beams of electron anti-neutrinos, and other neutrinos.  Such 
beams can be created with reasonable energy efficiency by a 
civilization without a long stretch from the technology we now 
possess.  Detection of such a beam, possible given a detecting 
civilization with our present level of neutrino detectors (cubic 
kilometer scale), would be evidently due to ETI with only a few events 
detected since there is no known mechanism for making neutrinos at 
only this energy range. Plus, having a few as two neutrinos arrive 
from precisely the same direction would be very unlikely unless 
accompanied by a huge burst of radiation in other bands. Data would 
accumulate at a rate plausibly about 1 Hz equivalent bandwidth and 
decoding the pattern would take perhaps one year.  This would amount 
to transmission of 1000 pages of material per year, a tremendous 
amount of information.  Given that the transmitting entity would have 
to know about the earth's (and other targets') ephemeris and possibly 
even day cycle, their transmission might be set to repeat several 
times daily and again on a longer cycle. Given that the ETI have no a 
priori knowledge of when observations will begin, and cannot get 
feedback for millenia afterward, multiple transmissions would be 
necessary, but perhaps only once in a few years, as perhaps they would 
illuminate other systems alternately and if our picture of the device 
is at all accurate, redirecting the accelerator would require some 
time.

No special action is required on our part, since if this speculative 
hypothesis should be correct, we will soon discover such signatures, 
but perhaps such will not arrive for some time. This adds motivation 
to keep all neutrino telescopes operating for long timescales, such as 
the watch for supernovae in our galaxy and unpredictable burst events 
of all types.  We humans should certainly think about and continue to 
explore other means for such communications, but to us neutrinos seem 
to provide some special opportunities.

\begin{acknowledgments}

This work was supported in part by the U.S.D.O.E. under grant 
DE-FG02-04ER41291 at University of Hawaii, and by the N.S.F. under 
grant 04-56556 at University of California at Santa Barbara. Two of us 
(S.P. and A.Z.) would also like to thank Xiao-Gang He, Pauchy Hwang 
and their colleagues at the NTU for their hospitality and the 
stimulating atmosphere where this work was begun.  We also thank 
Xerxes Tata for useful discussions.

\end{acknowledgments}


\end{document}